\documentstyle[12pt]{article}
\def \bt{\bar T}
\def \bc{\bar C}
\def \be{\bar E}

\def \ks{K^*,K^{**}}
\def \ra{\rho,a_2}
\def \tt{\tilde T}
\def \tc{\tilde C}
\def \te{\tilde E}
\def \ta{\tilde A}
\def\that{{\hat T}}

\def\phat{{\hat P}}
\def\ehat{{\hat E}}

\newcommand{\thspace}{\kern.08333em}

\def \beq{\begin{equation}}
\def \eeq{\end{equation}}
\def \beqn{\begin{eqnarray}}
\def \eeqn{\end{eqnarray}}

\def \s{\sqrt{2}}

\def \v#1#2{V_{#1#2}}

\begin{document}
\rightline{EFI-97-04}
\rightline{TECHNION-PH-97-3}
\rightline{January 1997}
\bigskip
\bigskip
\centerline{{\bf Annihilation, Rescattering, and CP Asymmetries in $B$ Meson
Decays}
\footnote{To be submitted to Phys.~Rev.~Lett.}}
\bigskip
\centerline{\it Boris Blok and Michael Gronau}
\centerline{\it Department of Physics}
\centerline{\it Technion -- Israel Institute of Technology, Haifa 32000,
Israel}
\medskip
\centerline{and}
\medskip
\centerline{\it Jonathan L. Rosner}
\medskip
\centerline{\it Enrico Fermi Institute and Department of Physics}
\centerline{\it University of Chicago, Chicago, IL 60637}
\bigskip
\centerline{\bf ABSTRACT}
\medskip
\begin{quote}
A number of $B$ meson decays may proceed only through participation of the
spectator quark, whether through amplitudes proportional to $f_B/m_B$ or via
rescattering from other less-suppressed amplitudes.  An expected hierarchy of
amplitudes in the absence of rescattering will be violated by rescattering
corrections.  Such violations could point the way toward channels in which
final-state interactions could be important. Cases in which final state phases
can lead to large CP asymmetries are pointed out.

\end{quote}
\medskip
\leftline{\qquad PACS codes:  12.15.Hh, 12.15.Ji, 13.25.Hw, 14.40.Nd}
\newpage

The decays of $B$ mesons have the potential for exhibiting CP violation under a
variety of conditions \cite{BCP}. Decays to CP eigenstates like $J/\psi K_S$
and $\pi^+ \pi^-$ are expected to display an appreciable time-dependent rate
asymmetry between $B^0$ and $\bar B^0$, whose interpretation in terms of pure
CKM phases relies on the assumption of the dominance of a single weak phase
\cite{PEN}.
Decays to non-CP eigenstates also can exhibit rate asymmetries in the presence
of at least two contributing amplitudes whose weak and strong phases both
differ from one another. However, the strong phases cannot be evaluated
reliably {\it a priori}. Instead, one must rely on constructions based on
amplitude triangles or quadrangles \cite{triquad}, in which one can separate out
weak from strong phases by comparing rates for processes with those for their
charge-conjugates.

In the present note we propose a test for large final-state interactions which,
while it does not yield precisely quantitative information on final-state
phases, can indicate in which channels such phases are likely to be large.
These channels are then prospects for searches for CP violation in rate
asymmetries. We will only discuss direct CP asymmetries between instantaneous
decays of $B$ mesons of opposite flavors, disregarding (in the case of neutral
$B$ mesons) time-dependent $B-\bar B$ mixing effects. Such asymmetries, which
require flavor-tagging, are obtained by time-integrated measurements and can be
carried out also in a symmetric $e^+e^-$ collider operating at the
$\Upsilon(4S)$.

We consider processes $B \to P P$, where $P$ is a pseudoscalar meson. Similar
results hold when one or both pseudoscalars are replaced by vector mesons $V$.
There are many processes in which the spectator quark necessarily plays a role
in the decay, whether through exchange ($E$) or annihilation ($A$) with the $b$
quark or via rescattering.  In the case of exchange or annihilation, the decay
amplitude is expected to contain a power of $f_B/m_B$, where $f_B \simeq 200$
MeV is the $B$ meson decay constant. Such an amplitude is expected to be
suppressed by a factor of roughly $\lambda^2$ \cite{hier}, where $\lambda=0.22$
is a parameter introduced by Wolfenstein \cite{WP} to classify the
hierarchy of elements in the Cabibbo-Kobayashi-Maskawa (CKM) matrix.

Processes in which the $b$ quark decay contributes to final states without the
intervention of the spectator quark are expected to dominate $B \to P P$
decays. These processes consist of tree amplitudes which are color-favored
($T$) or color-suppressed ($C$) and penguin ($P$) amplitudes. (The notation is
that of \cite{hier}.)
In many cases such amplitudes can contribute through rescattering to the
processes involving the participation of the spectator quark \cite{D}.
Our purpose is to enumerate and classify these situations.

In Table I we list all the $B \to P P$ processes for which $T$, $C$, or $P$
amplitudes cannot contribute except via rescattering.  Also shown are the
powers of $\lambda$ in the amplitudes (which are the same whether the
amplitudes arise from exchange or annihilation or via rescattering from $T$ or
$C$). Any process of order $\lambda^n$ should effectively appear of order
$\lambda^n (f_B/m_B) \simeq {\cal O}(\lambda^{n+2})$ if rescattering is not
important, whereas rescattering could in principle enhance this amplitude
beyond this level to a maximum of ${\cal O} (\lambda^n)$.

The most promising decays have amplitudes of order $f_B \lambda^2/m_B$. It
would be interesting to search for the modes in Table I to see if they show
evidence for the expected suppression due to the factor $f_B/m_B$. If they are
enhanced, there is some chance that the rescattering amplitude can generate a
final-state phase large enough to give rise to an observable CP-violating rate
asymmetry in the decay. In the following discussion we will describe a possible
mechanism for such enhancement.

Recently some progress was made in understanding the role of final state
interactions in $B$ decays \cite{DESH,ZH,DGPS,BH,NP}. In Ref.~\cite{DGPS}
it was argued that, contrary to simple intuition \cite{BJ}, soft final state
interactions do not disappear in
the large $m_B$ limit and may be significant in hadronic $B$ decays.
In Ref.~\cite{BH} sizable rescattering effects via inelastic
charge-exchange,
\beq
\pi^+ D^-\to \pi^0 \bar D^0 ~,
\eeq
were calculated in $B^0 \to \pi^0 \bar D^0$. Let us summarize the results of
this analysis, which will then be applied to our case.

The processes $B^0 \to \pi^+ D^-$ and $B^0 \to \pi^0 \bar D^0$ are
conventionally given by ``color-allowed" and ``color-suppressed" amplitudes,
$\bt \propto a_1$ and $\bc \propto a_2$, respectively, which are determined
experimentally, $a_2/a_1 \approx 0.2$ \cite{CLEO}. When calculated in the
na\"{\i}ve factorization approximation \cite{BSW}, neglecting rescattering
effects, both amplitudes are real. The new contribution to $B^0 \to \pi^0 \bar
D^0 $ decay, via the rescattering process (1), was calculated in terms of $\rho$
trajectory Regge exchange. Denoting the direct decay amplitude
$A(B^0 \to \pi^0
\bar D^0) $ by $M^{dir}_{\pi D}$ and the decay amplitude to the same final
state via charge-exchange, $A(B^0 \to \pi^+ D^-\to \pi^0 \bar D^0)$, by
$M^r_{\pi D}$, it was found that \cite{BH}
\beq
\label{DPI}
{M^{r}_{\pi D} \over M^{dir}_{\pi D}} \approx 0.18 + 0.58i~.
\eeq
That is, the additional contribution from rescattering into the $\pi^0 \bar D^0$
final state is as important as the direct amplitude and is dominantly imaginary.
Thus, the total amplitude of $B^0 \to \pi^0 \bar D^0$, given by $M^{r}_{\pi D} +
M^{dir}_{\pi D}$, is predicted to carry a large final state phase.
A similar situation exists in $B^0 \to \pi^0\pi^0$ due to rescattering effects
from $B^0 \to \pi^+ \pi^-\to \pi^0 \pi^0$. We wish to stress that the
calculation leading to Eq.~(2) involves quite a few assumptions, and can
therefore mainly serve for illustrative purposes \cite{PIPI}. Smaller
rescattering effects were calculated in Ref.~\cite{NP} for proceses of the type
$B\to PV$.

To sum up, the results of the analysis of Ref.~\cite{BH} suggest that in the
two cases of $B^0 \to \pi^0 \bar D^0$ and $B^0 \to \pi^0\pi^0$, the
rescattering amplitudes into the final states may carry large phases, and are
likely to be smaller by a factor of about $\lambda$ than the decay amplitudes
to the intermediate $\pi^+ D^-$ and $\pi^+\pi^-$ states, respectively. The
existence of large phases can be tested experimentally by measuring the rates
of $B^0 \to \pi^0 \bar D^0$ and $B^0 \to \pi^0\pi^0$ and of isospin-related
processes. Since the final states are mixtures of two isospin states
($I=1/2,~3/2$ and $I=0,~2$, respectively), the amplitudes of these processes
obey two triangle relations with the amplitudes of two other pairs of
processes, $B^0 \to \pi^+ D^-,~B^+ \to \pi^+ \bar D^0$ and $B^0 \to \pi^+
\pi^-,~ B^+ \to \pi^+ \pi^0$, respectively. Rescattering effects into these
final states are expected to be smaller than in the color-suppressed processes.
Thus, the smaller sides of the two triangles, associated with the $B^0 \to
\pi^0 \bar D^0$ and $B^0 \to \pi^0\pi^0$ amplitudes, will form sizable angles
with each of the other two sides. In the case of the $B \to \pi \pi$ isospin
triangle, one would have to isolate the contribution from a penguin amplitude
which carries a different {\em weak} phase \cite{LON}. So far, the two large
sides of the $B\to \pi \bar D$ triangle have been measured, while an upper
limit exists on $B^0 \to \pi^0 \bar D^0$. This sets a mild upper bound on the
corresponding phase \cite{YAMA}.

A similar situation is expected to hold in the processes on the left-hand-side
of Table 1. In addition to the direct $E$ and $A$ amplitudes, which are
suppressed by $f_B/m_B$, these processes obtain contributions of rescattering
from intermediate states given in the right-hand-side of the Table. Also
specified are the types ($T, C, P$) of decay amplitudes into the intermediate
states and the corresponding Regge trajectories ($K^*, K^{**}, \rho, a_2$)
exchanged between the intermediate and final states.

As an example, consider $B^0 \to K^+D^-_s$. The amplitude of this process
contains two terms: the direct $\be$ amplitude and the contributions of
rescattering from $\pi^+D^-$ and from $\pi^0 \bar D^0$ states, which are
described in terms of $K^*, K^{**}$ Regge exchange. The decay amplitudes into
$\pi^+D^-$ ($\bt$) dominates over the decay amplitude into $\pi^0 \bar D^0$
($\bc/\s$). In the chiral limit (in which the mass of the $s$ quark vanishes,
$m_s\rightarrow 0$), $K^*$ exchange is equivalent to $\rho$ exchange. Chiral
corrections and the $K^{**}$ trajectory are expected to increase the amplitude.
The contribution to $B^0 \to K^+D^-_s$, via the rescattering process $\pi^+D^-
\to K^+D^-_s$ is approximately equal to the contribution to $B^0 \to \pi^0 \bar
D^0$ via the rescattering process (1). Thus, assuming the results of
Ref.~\cite{BH}, we find that the amplitude of $B^0 \to K^+D^-_s$ obtains two
terms: a direct amplitude $\be$ which is real and of order $\lambda^4$, and a
contribution from rescattering via $\pi^+ D^-$, which is of order $\lambda^3$
and which carries a large final state phase. Thus, $A(B^0 \to K^+D^-_s)$ is
expected to be enhanced by a factor $1/\lambda$ compared to the na\"{\i}ve
$f_B/m_B$ suppression and to obtain a sizable final-state interaction phase.
The enhancement can be tested by measuring the rate of this process.

Similar effects exist in all the other processes in Table 1. For instance, the
amplitude of $B^+ \to K^0 D^+$ consists of a real direct term of order
$\lambda^5$, and an amplitude due to rescattering from $\pi^0 D^+_s$ which is
of order $\lambda^4$ and has a large strong phase.

We conclude that rescattering from intermediate states leads to amplitudes
suppressed by $\lambda$ rather than by $f_B/m_B$. The presence of large final
state phases in these amplitudes does not guarantee large CP asymmetries. For
this one requires that two different weak phases contribute to a process. In
$B^0 \to K^+D^-_s$, $\bt$ ($\bc$) and $\be$ involve the same weak phase ${\rm
Arg}(V^*_{cb}V_{ud})$, and no CP asymmetry is expected between the rate of this
process and its charge-conjugate. A similar situation holds in $B^+ \to K^0
D^+$.

In order to search for cases in which CP asymmetries can be expected
as a consequence of two different weak phases, we limit our attention to
processes in Table 1, in which the contribution to an amplitude from
rescattering involves a penguin ($P$) term. The weak phase of this amplitude
may differ from the phase of the direct $E$ or $A$ amplitude. There are four
such classes of processes.

In the first class, $B_s \to D^+D^-$ and $B_s \to D^0 \bar D^0$, all amplitudes
involve the same CKM phase, ${\rm Arg}(V^*_{cb}V_{cs}) = {\rm Arg}
(V^*_{tb}V_{ts})~({\rm mod}~\pi)$, and one expects no CP asymmetry.

In the second and third class, involving $B^0 \to D^0 \bar D^0$ or $D^+_s
D^-_s$ and $B^0 \to K^+ K^-$, respectively, the direct amplitude and the penguin
contribution to rescattering involve different weak phases, ${\rm Arg}
(V^*_{cb}V_{cd}) \ne {\rm Arg}(V^*_{tb}V_{td}),~{\rm Arg}(V^*_{ub}V_{ud}) \ne
{\rm Arg}(V^*_{tb}V_{td})$. (We neglect the effect of $u$ and $c$ quarks in $b
\to d$ penguin amplitudes \cite{BF}). The asymmetries in the processes
belonging to these two classes are proprtional to the sines of the
corresponding weak phase differences, namely to $\sin\beta$ and $\sin\alpha$,
respectively, where $\alpha$ and $\beta$ are two angles of the CKM unitarity
triangle \cite{BCP}. Since the penguin amplitude is subdominant in the decays
to intermediate states, the rescattering effects in the asymmetries are
suppressed by
\beq
\vert{\phat'\over \that'}\vert \sim \vert{V^*_{tb}V_{td}\over
V^*_{cb}V_{cd}}\vert
{\alpha_s(m_b)\over 12\pi}\ln({m^2_t\over m^2_b}) \sim \lambda ~{\rm to}
~\lambda^2~,
\eeq
and by
\beq
\vert{P\over T}\vert \sim \vert{V^*_{tb}V_{td}\over V^*_{ub}V_{ud}}\vert
{\alpha_s(m_b)\over 12\pi}\ln({m^2_t\over m^2_b}) \sim \lambda ~{\rm to}
~\lambda^2~,
\eeq
respectively. Consequently, CP asymmetries are estimated at the level of 10\% in
$B^0 \to D^0 \bar D^0$ or $D^+_s D^-_s$ and in $B^0 \to K^+ K^-$.

The largest asymmetries are expected in $B_s \to \pi^+\pi^-$ and $B_s \to
\pi^0\pi^0$. In these rather rare processes, the direct amplitudes are of order
$\lambda^6$ and carry a weak phase ${\rm Arg}(V^*_{ub}V_{us})$. The penguin
amplitude $P'$ dominates $B_s$ decays to the $K \bar K$ intermediate states.
Its weak phase is ${\rm Arg}(V^*_{tb}V_{ts})$, and its magnitude can be
estimated by
\beq
P' \sim \vert V^*_{tb}V_{ts}\vert (\alpha_s(m_b)/12\pi)\ln(m^2_t/m^2_b)
\sim \lambda^4~.
\eeq
The rescattering amplitudes into the charged and neutral $\pi\pi$ states are
smaller by a factor $\lambda$ and carry large final state phases. Thus, the
magnitudes of the two interfering amplitudes differ by no more than one power
of $\lambda$, their weak phase-difference is $\gamma$ and their strong phase
difference is likely to be large. One therefore expects in $B_s \to \pi^+\pi^-$
and in $B_s \to \pi^0\pi^0$ large asymmetries, possibly of a few tens of
percent, which are proportional to $\sin\gamma$.

Let us note that similar rescattering effects inducing CP asymmetries
are expected also in processes which do not require participation of the
spectator quark. One such example, in which large asymmetries are expected, is
$B^0 \to K^0 \bar K^0$. In this case, the direct amplitude
is pure penguin \cite{hier} and has a magnitude
\beq
P \sim \vert V^*_{tb}V_{td}\vert (\alpha_s(m_b)/12\pi)\ln(m^2_t/m^2_b)
\sim \lambda^4~{\rm to}~\lambda^5~.
\eeq
The $K^0 \bar K^0$ final state can be also obtained by decay to and
rescattering from a $\pi^+\pi^-$ state.  The amplitude of $B^0 \to \pi^+\pi^-$
is given by $T \sim V^*_{ub}V_{ud} \sim\lambda^3~{\rm to}~\lambda^4$; the
rescattering amplitude into $K^0 \bar K^0$ is smaller by a factor $\lambda$ and
carries a large strong phase. Hence, the two amplitudes are of comparable
magnitudes and may have quite different strong phases. The resulting
asymmetry which is proportional to $\sin\alpha$ can be sizable. A similar
result was obtained in
Ref.~\cite{GH}, in which the rescattering amplitude via the $\pi^+\pi^-$ state
was represented by the soft part of the $u$ quark contribution to the penguin
amplitude $P$.

A remark is in order about the magnitude of the factor $f_B/m_B \simeq
\lambda^2$ which we have assumed to characterize suppressed amplitudes. Such a
suppression of exchange and annihilation amplitudes of $B$ decays to two
pseudoscalars is obtained when assuming factorization and simple form factors
\cite{XING}. This picture is clearly an oversimplification. As we have shown,
it is quite possible that these amplitudes may be suppressed only by $\lambda$
due to rescattering effects. A similar suppression, of order $2 \pi f_B/m_B
\simeq \lambda$, characterizes the inclusive annihilation amplitude \cite{VOL}.
The extra $2\pi$ factor has a simple explanation. The non-suppressed amplitudes
involve beta-decay type transitions in which one quark decays to three, whereas
the suppressed amplitudes involve exchange or annihilation of an initial heavy
quark and an initial spectator antiquark into a {\em two}-quark final state.
The factor of $2 \pi$ takes account of the presence of one less quark in the
final state.

We comment briefly on expected branching ratios for the interesting processes.
One must distinguish between those processes in which the magnitude of
rescattering is expected to be merely detectable and those in which it can lead
to a measurable CP-violating decay rate asymmetry.  For example, the amplitude
for $B^0 \to K^+ D_s^-$ is expected to be of order $\lambda$ times that for the
observed process $B^0 \to \pi^+ D^-$ which feeds it via rescattering.  Since
the branching ratio for the latter process is about 0.3\%, observation of a
rate for $B^0 \to K^+ D_s^-$ high enough to imply rescattering effects, namely
with a branching ratio of about $10^{-4}$, is within the reach of present or
modestly upgraded $B$ meson experiments.

The use of the anticipated large final state phases to observe a CP-violating
asymmetry is somewhat more demanding, but within the capabilities of several
planned high-intensity sources of $B$ mesons. The decays $B^0 \to D^0 \bar D^0$
and  $B^0 \to D^+_s D^-_s$ are expected to be fed by rescattering from both
tree and penguin contributions in $B^0 \to D^+ D^-$. One expects the branching
ratio for the latter process to be about $\lambda^2$ that for the observed
process $B^0 \to D^- D_s^+$, or about $(1/20)(0.8\%)$. Rescattering will
probably cost another factor of $\lambda^2$ in rate, leading to a branching
ratio $B(B^0 \to D^0 \bar D^0) \simeq 2\times 10^{-5}$. As we have estimated, a
rate asymmetry of order 10\% could arise between this process and its
charge-conjugate. The situation in $B^0 \to K^+ K^-$, where an asymmetry at a
similar level is expected, is somewhat less favorable. This process is fed by
rescattering from $B^0 \to \pi^+ \pi^-$, the rate of which is likely to be
about $10^{-5}$ \cite{CLEOpp}, so $B(B^0 \to K^+ K^-)$ should be somewhat below
$10^{-6}$. Finally, rates at a similar level are expected for $B_s \to
\pi^+\pi^-$ and $B_s \to \pi^0\pi^0$, which are fed by rescattering from $B_s
\to K^+K^- (K^0 \bar K^0)$ whose branching ratios is probably about $10^{-5}$,
similar to that of $B^0 \to K^+ \pi^-$ \cite{CLEOpp}. As we noted, the asymmetry
in $B_s \to\pi\pi$ may be very large due to the interference between the
amplitudes $E'$ and the rescattering amplitude from $P'$ which differ by
no more than one power of $\lambda$.

In summary, while it is very difficult to study quantitatively final state
interactions at the $B$ mass, our analysis indicates that in $B$ decays
which require the participation of the spectator quark, rescattering effects
are likely to enhance decay rates by an order of magnitude relative to
na\"{\i}ve expectations. Such enhancement may indicate large final state
phases, which would lead in certain cases to sizable CP asymmetries.
\bigskip

We thank J. Alexander, J. Bartelt, G. C. Moneti, D. Wyler, and D-X. Zhang
for discussions and the CERN Theory Group for a congenial atmosphere in which
part of this collaboration was carried out. This work was supported in part by
the United States -- Israel Binational Science Foundation under Research Grant
Agreement 94-00253/2, by the Israel Science Foundation, and by the United
States Department of Energy under Contract No. DE FG02 90ER40560.
\newpage

\def \ajp#1#2#3{Am.~J.~Phys.~{\bf#1}, #2 (#3)}
\def \apny#1#2#3{Ann.~Phys.~(N.Y.) {\bf#1}, #2 (#3)}
\def \app#1#2#3{Acta Phys.~Polonica {\bf#1}, #2 (#3)}
\def \arnps#1#2#3{Ann.~Rev.~Nucl.~Part.~Sci.~{\bf#1}, #2 (#3)}
\def \cmp#1#2#3{Commun.~Math.~Phys.~{\bf#1}, #2 (#3)}
\def \cmts#1#2#3{Comments on Nucl.~Part.~Phys.~{\bf#1}, #2 (#3)}
\def \corn93{{\it Lepton and Photon Interactions:  XVI International
Symposium, Ithaca, NY August 1993}, AIP Conference Proceedings No.~302,
ed.~by P. Drell and D. Rubin (AIP, New York, 1994)}
\def \cp89{{\it CP Violation,} edited by C. Jarlskog (World Scientific,
Singapore, 1989)}
\def \dpff{{\it The Fermilab Meeting -- DPF 92} (7th Meeting of the
American Physical Society Division of Particles and Fields), 10--14
November 1992, ed. by C. H. Albright \ite~(World Scientific, Singapore,
1993)}
\def \dpf94{DPF 94 Meeting, Albuquerque, NM, Aug.~2--6, 1994}
\def \efi{Enrico Fermi Institute Report No. EFI}
\def \el#1#2#3{Europhys.~Lett.~{\bf#1}, #2 (#3)}
\def \f79{{\it Proceedings of the 1979 International Symposium on Lepton
and Photon Interactions at High Energies,} Fermilab, August 23-29, 1979,
ed.~by T. B. W. Kirk and H. D. I. Abarbanel (Fermi National Accelerator
Laboratory, Batavia, IL, 1979}
\def \hb87{{\it Proceeding of the 1987 International Symposium on Lepton
and Photon Interactions at High Energies,} Hamburg, 1987, ed.~by W. Bartel
and R. R\"uckl (Nucl. Phys. B, Proc. Suppl., vol. 3) (North-Holland,
Amsterdam, 1988)}
\def \ib{{\it ibid.}~}
\def \ibj#1#2#3{~{\bf#1}, #2 (#3)}
\def \ichep72{{\it Proceedings of the XVI International Conference on High
Energy Physics}, Chicago and Batavia, Illinois, Sept. 6--13, 1972,
edited by J. D. Jackson, A. Roberts, and R. Donaldson (Fermilab, Batavia,
IL, 1972)}
\def \ijmpa#1#2#3{Int.~J.~Mod.~Phys.~A {\bf#1}, #2 (#3)}
\def \ite{{\it et al.}}
\def \jmp#1#2#3{J.~Math.~Phys.~{\bf#1}, #2 (#3)}
\def \jpg#1#2#3{J.~Phys.~G {\bf#1}, #2 (#3)}
\def \lkl87{{\it Selected Topics in Electroweak Interactions} (Proceedings
of the Second Lake Louise Institute on New Frontiers in Particle Physics,
15--21 February, 1987), edited by J. M. Cameron \ite~(World Scientific,
Singapore, 1987)}
\def \ky85{{\it Proceedings of the International Symposium on Lepton and
Photon Interactions at High Energy,} Kyoto, Aug.~19-24, 1985, edited by M.
Konuma and K. Takahashi (Kyoto Univ., Kyoto, 1985)}
\def \mpla#1#2#3{Mod.~Phys.~Lett.~A {\bf#1}, #2 (#3)}
\def \nc#1#2#3{Nuovo Cim.~{\bf#1}, #2 (#3)}
\def \np#1#2#3{Nucl.~Phys.~{\bf#1}, #2 (#3)}
\def \pisma#1#2#3#4{Pis'ma Zh.~Eksp.~Teor.~Fiz.~{\bf#1}, #2 (#3) [JETP
Lett. {\bf#1}, #4 (#3)]}
\def \pl#1#2#3{Phys.~Lett.~{\bf#1}, #2 (#3)}
\def \plb#1#2#3{Phys.~Lett.~B {\bf#1}, #2 (#3)}
\def \pr#1#2#3{Phys.~Rev.~{\bf#1}, #2 (#3)}
\def \pra#1#2#3{Phys.~Rev.~A {\bf#1}, #2 (#3)}
\def \prd#1#2#3{Phys.~Rev.~D {\bf#1}, #2 (#3)}
\def \prl#1#2#3{Phys.~Rev.~Lett.~{\bf#1}, #2 (#3)}
\def \prp#1#2#3{Phys.~Rep.~{\bf#1}, #2 (#3)}
\def \ptp#1#2#3{Prog.~Theor.~Phys.~{\bf#1}, #2 (#3)}
\def \rmp#1#2#3{Rev.~Mod.~Phys.~{\bf#1}, #2 (#3)}
\def \rp#1{~~~~~\ldots\ldots{\rm rp~}{#1}~~~~~}
\def \si90{25th International Conference on High Energy Physics, Singapore,
Aug. 2-8, 1990}
\def \slc87{{\it Proceedings of the Salt Lake City Meeting} (Division of
Particles and Fields, American Physical Society, Salt Lake City, Utah,
1987), ed.~by C. DeTar and J. S. Ball (World Scientific, Singapore, 1987)}
\def \slac89{{\it Proceedings of the XIVth International Symposium on
Lepton and Photon Interactions,} Stanford, California, 1989, edited by M.
Riordan (World Scientific, Singapore, 1990)}
\def \smass82{{\it Proceedings of the 1982 DPF Summer Study on Elementary
Particle Physics and Future Facilities}, Snowmass, Colorado, edited by R.
Donaldson, R. Gustafson, and F. Paige (World Scientific, Singapore, 1982)}
\def \smass90{{\it Research Directions for the Decade} (Proceedings of the
1990 Summer Study on High Energy Physics, June 25 -- July 13, Snowmass,
Colorado), edited by E. L. Berger (World Scientific, Singapore, 1992)}
\def \stone{{\it B Decays}, edited by S. Stone (World Scientific,
Singapore, 1994)}
\def \tasi90{{\it Testing the Standard Model} (Proceedings of the 1990
Theoretical Advanced Study Institute in Elementary Particle Physics,
Boulder, Colorado, 3--27 June, 1990), edited by M. Cveti\v{c} and P.
Langacker (World Scientific, Singapore, 1991)}
\def \yaf#1#2#3#4{Yad.~Fiz.~{\bf#1}, #2 (#3) [Sov.~J.~Nucl.~Phys.~{\bf #1},
#4 (#3)]}
\def \zhetf#1#2#3#4#5#6{Zh.~Eksp.~Teor.~Fiz.~{\bf #1}, #2 (#3) [Sov.~Phys.
- JETP {\bf #4}, #5 (#6)]}
\def \zpc#1#2#3{Zeit.~Phys.~C {\bf#1}, #2 (#3)}

\newpage

\renewcommand{\arraystretch}{1.2}

\begin{table}
\caption{$B \to PP$ amplitudes involving only suppressed graphs.}
\begin{center}
\begin{tabular}{c c c c c c c} \hline
CKM  & Order   & Process & Suppressed & \multicolumn{3}{c}{Rescatters from:} \\
\cline{5-7}
Factor & in $\lambda$ &    & amplitude  & State & Ampl. & Exch. \\ \hline
$V_{cb}^*V_{ud}$ & $\lambda^2$ & $B^0 \to K^+ D_s^-$ & $\be$ &
  $\pi^+ D^-$ & $\bt$ & $\ks$ \\
                 &                 &                   &      &
  $\pi^0 \bar D^0$ & $\bc/\s$ & $\ks$ \\ \hline
$V_{cb}^*V_{cs}$ & $\lambda^2$ & $B_s \to D^+ D^-$ & $\ehat$ &
  $D_s^+ D_s^-$ & $\that + \phat$ & $\ks$ \\
                             & & $B_s \to D^0 \bar D^0$ & $-\ehat$ &
  $D_s^+ D_s^-$ & $\that + \phat$ & $\ks$ \\ \hline
$V_{cb}^*V_{us}$ & $\lambda^3$ & $B_s \to \pi^+ D^-$ & $\lambda \be$ &
  $K^+ D_s^-$ & $\lambda \bt$ & $\ks$ \\
                             & & $B_s \to \pi^0 \bar D^0$ & $-\lambda \be/\s$ &
  $K^+ D_s^-$ & $\lambda \bt$ & $\ks$ \\ \hline
$V_{cb}^*V_{cd}$ & $\lambda^3$ & $B^0 \to D^0 \bar D^0$ & $-\ehat'$ &
  $D^+ D^-$ & $\that' + \phat'$ & $\ra$ \\
                             & & $B^0 \to D_s^+ D_s^-$ & $\ehat'$ &
  $D^+ D^-$ & $\that' + \phat'$ & $\ks$ \\ \hline
$V_{ub}^*V_{ud}$ & $\lambda^{3}$ & $B^0 \to K^+ K^-$ & $-E$ &
  $\pi^+ \pi^-$ & $-(T+P)$ & $\ks$ \\
                 &                 &                   &      &
  $\pi^0 \pi^0$ & $(P-C)/\s$ & $\ks$ \\
                 &                 &                   &      &
  $K^0 \bar K^0$ & $P$ & $\ra$ \\ \hline
$V_{ub}^*V_{cs}$ & $\lambda^3$ & $B^+ \to K^0 D^+$ & $\ta$     &
  $\pi^0 D_s^+$ & $-\tt/\s$ & $\ks$ \\
                 &                 &                   &      &
  $K^+ D^0$ & $-\tc$ & $\ra$ \\
                             & & $B_s \to \pi^- D^+$ &  $-\te$ &
  $K^- D_s^+$ & $-\tt$ & $\ks$ \\
                             & & $B_s \to \pi^0 D^0$ & $\te/\s$ &
  $K^- D_s^+$ & $-\tt$ & $\ks$ \\ \hline
$V_{ub}^*V_{us}$ & $\lambda^{4~a}$ & $B_s \to \pi^+ \pi^-$ & $-E'$ &
  $K^+ K^-$ & $-(T'+P')$ & $\ks$ \\
                 &                 &                   &      &
  $K^0 \bar K^0$ & $P'$ & $\ks$ \\
                             & & $B_s \to \pi^0 \pi^0$ & $E'/\s$ &
  $K^+ K^-$ & $-(T'+P')$ & $\ks$ \\
                 &                 &                   &      &
  $K^0 \bar K^0$ & $P'$ & $\ks$ \\ \hline
$V_{ub}^*V_{cd}$ & $\lambda^4$ & $B^+ \to \bar K^0 D_s^+$ & $-\lambda\ta$ &
  $\pi^+ D^0$ & $\lambda \tc$ & $\ks$ \\
                 &                 &                   &      &
  $\pi^0 D^+$ & $\lambda \tt/\s$ & $\ks$ \\
                             & & $B^0 \to K^- D_s^+$ & $\lambda\te$ &
  $\pi^- D^+$ & $\lambda \tt$ & $\ks$ \\
                 &                 &                   &      &
  $\pi^0 D^0$ & $\lambda \tc/\s$ & $\ks$ \\ \hline
\end{tabular}
\end{center}
\leftline{$^a$Penguin annihilation (also of this order) ignored}
\end{table}

\begin{thebibliography}{99}
\bibitem{PIPI} In the calculation of Ref.~\cite{BH} the phases of the two
rescattered amplitudes $A(B^0 \to \pi^+ D^- \to \pi^0 D^0)$ and
$A(B^0 \to \pi^+ \pi^- \to \pi^0 \pi^0)$ are the sums of two large unrelated
phases which involve large theoretical uncertainties. In both cases the
magnitude of $M^{r}/ M^{dir}$ is approximately one.
\end{thebibliography}
\end{document}